\documentclass[aps,prc,12pt,showpacs,showkeys,superscriptaddress,groupedaddress]{revtex4-1}

\usepackage[english]{babel}
\usepackage{indentfirst}
\usepackage{amssymb}
\usepackage{braket}
\usepackage{bm}
\usepackage{amsxtra}
\usepackage{amsmath}
\usepackage{supertabular}
\usepackage{multirow}
\usepackage{color}
\usepackage [mathcal]{eucal}
\usepackage{graphicx}
\usepackage{graphics}
\usepackage{lipsum}

\def\n4lo{$\mathrm{N}^4\mathrm{LO}$}
\def\n3lo{$\mathrm{N}^3\mathrm{LO}$}
\def\dblone{\hbox{$1\hskip -1.2pt\vrule depth 0pt height 1.6ex width 0.7pt
     \vrule depth 0pt height 0.3pt width 0.12em$}}
     \def\chiral4lo{$\mathrm{N}^4\mathrm{LO}$}

\begin{document}

\noindent
\title{Optical Potentials Derived from 
Nucleon-Nucleon Chiral Potentials at N$^4$LO}

\author{Matteo Vorabbi$^{1}$}
\author{Paolo Finelli$^{2}$}
\author{Carlotta Giusti$^{3}$}

\affiliation{$~^{1}$TRIUMF, 4004 Wesbrook Mall, Vancouver, British Columbia, V6T 2A3, Canada
}

\affiliation{$~^{2}$Dipartimento di Fisica e Astronomia, 
Universit\`{a} degli Studi di Bologna and \\
INFN, Sezione di Bologna, Via Irnerio 46, I-40126 Bologna, Italy
}

\affiliation{$~^{3}$Dipartimento di Fisica,  
Universit\`a degli Studi di Pavia and \\
INFN, Sezione di Pavia,  Via A. Bassi 6, I-27100 Pavia, Italy
}

\date{\today}


\begin{abstract} 

{\bf Background:} Elastic scattering is probably the main event in the interactions of nucleons with nuclei. Even if this process has 
been extensively studied in the last years, a consistent description, i.e., starting from microscopic two- and many-body forces connected by the same symmetries and principles, is still under development.

{\bf Purpose:} In a previous paper~\cite{PhysRevC.93.034619} we derived a theoretical optical potential from $NN$ chiral potentials at fourth order (N$^3$LO).
 
In the present work we use $NN$ chiral potentials at fifth order (N$^4$LO), with the purpose to check the convergence and to assess the theoretical errors associated with the truncation of the chiral expansion in the construction of an optical potential.

{\bf Methods:} Within the same framework and with the same approximations as Ref.~\cite{PhysRevC.93.034619}, the optical potential is derived as the first-order term within the spectator expansion of the nonrelativistic multiple scattering theory and adopting the impulse approximation and the optimum factorization approximation.

{\bf Results:} The {\it pp} and {\it np} Wolfenstein amplitudes and the cross section, analyzing power, and spin rotation of elastic proton scattering from $^{16}$O, $^{12}$C, and $^{40}$Ca nuclei are presented at an incident proton energy of 200 MeV. The results obtained with different versions of chiral potentials at N$^4$LO are compared.

{\bf Conclusions:} Our results indicate that convergence has been reached at N$^4$LO. The agreement with the experimental data is comparable with the agreement obtained in Ref.~\cite{PhysRevC.93.034619}. We confirm that building an optical potential within chiral perturbation theory is a promising approach for describing elastic proton-nucleus scattering.

\end{abstract}

\pacs{24.10.-i; 24.10.Ht; 24.70.+s; 25.40.Cm}

\maketitle



\section{Introduction}
\label{sec_intro}

Elastic scattering is probably the main event in the interaction of nucleons with nuclei. A wealth of detailed information on nuclear properties has been obtained from the existing measurements of cross sections and polarization observables for the elastic scattering of protons from a wide variety of stable nuclei over a wide range of energies.
A suitable and successful framework to describe elastic  nucleon-nucleus ($NA$) scattering is provided by the nuclear optical potential~\cite{hodgson1963}. With the optical potential it is possible to compute the scattering observables  across wide regions of the nuclear landscape and to extend calculations to inelastic scattering and to a wide variety of nuclear reactions.

The optical potential can be derived phenomenologically or, alternatively and more fundamentally, microscopically.
Phenomenological optical potentials are obtained assuming a form and a dependence on a number of adjustable parameters for the real and the imaginary parts that characterize the shape of the nuclear density distribution and that vary with the nuclear energy and the nuclear mass number. The parameters are obtained through a fit to data of elastic proton-nucleus ($pA$) scattering data.
The calculation of a microscopic optical potential requires, in principle, the solution of the full many-body nuclear problem for the incident nucleon and the $A$ nucleons of the target, which is beyond present capabilities. In practice, with suitable approximations, microscopic optical potentials are usually derived from two basic quantities: the nucleon-nucleon ($NN$) $t$ matrix and the matter distribution of the nucleus. 

The $NN$ potential is an essential ingredient in the $NA$ scattering theory where its off-shell properties play an important role. To obtain a good description of these properties microscopic optical potentials are usually derived employing``realistic" $NN$ potentials, which are able to reproduce the experimental $NN$ phase shifts with a $\chi^2$/datum
$\simeq 1$.

In a previous paper of ours~\cite{PhysRevC.93.034619} a new microscopic optical potential for elastic $pA$ scattering has been obtained employing microscopic two-body chiral potentials, i.e., $NN$ potentials derived from first principles. The purpose of our work was just to study the domain of applicablity of chiral potentials in the construction of an optical potential. 
The theoretical framework basically follows the approach 
of Ref.~\cite{Kerman1959551}, where the Watson multiple scattering theory was developed expressing the
$NA$ optical potential by a series expansion in terms of the free $NN$ scattering amplitudes. In the calculations of Ref.~\cite{PhysRevC.93.034619} the expansion is  truncated at the first-order term, medium effects are neglected in the interaction between the projectile and the target nucleon and in the impulse approximation the interaction is described by the free $NN$ $t$ matrix. In addition, the optimum factorization approximation is adopted, where the optical potential is given by the factorized product of the free $NN$ $t$ matrix and the nuclear density.
For the $NN$ interaction, in Ref.~\cite{PhysRevC.93.034619} two different versions of chiral potentials at fourth order (N$^3$LO) in the chiral expansion have been used, developed by Entem and Machleidt~\cite{chiralmachleidt_n3lo} and Epelbaum, Gl\"ockle, and Mei\ss ner~\cite{chiralepelbaum_n3lo}.
The results produced by the two different versions of the chiral potential have been compared for the $NN$ scattering amplitudes and for the observables of elastic proton scattering on $^{16}$O. 

Recently, $NN$ potentials at fifth order (N$^4$LO) of chiral effective field theory have been presented by Epelbaum, Krebs, and Mei\ss ner (EKM)~\cite{Epelbaum:2014sza,Epelbaum:2014efa} and Entem, Machleidt, and 
Nosyk (EMN)~\cite{Entem:2014msa, Entem:2017gor}. These new chiral $NN$ potentials are used in the present work to calculate the optical potential within the same theoretical framework as in Ref.~\cite{PhysRevC.93.034619}.
The main aims of our work are to check the convergence of the Chiral Perturbation Theory (ChPT) expansion, to investigate the sensitivity of the results to the choice of the $NN$ potential and to the adopted regularization prescription, and to assess theoretical uncertainties on elastic $NA$ scattering observables.  

The paper is organized as follows: in Section~\ref{sec_mod} we outline the theoretical framework used to calculate the $NA$ optical 
potential. In Section~\ref{sec_pot} we introduce the chiral $NN$ potentials at fifth order recently presented in Refs.~\cite{Epelbaum:2014sza,Epelbaum:2014efa} (EKM) and~\cite{Entem:2014msa, Entem:2017gor} (EMN).
In Section~\ref{sec_res} we show and discuss our results for the $NN$ Wolfenstein amplitudes and for the scattering observables 
on a small set of light nuclei ($^{12}$C, $^{16}$O, and $^{40}$Ca) calculated with both $NN$ potentials. Predictions based on EKM  and EMN potentials are compared with available experimental data. 
Finally, in Section~\ref{sec_concl} we draw our conclusions.

\section{OPTICAL POTENTIALS}
\subsection{THEORETICAL FRAMEWORK}
\label{sec_mod}

Proton elastic scattering off a target nucleus with $A$ nucleons can be formulated in the 
momentum space by the full Lippmann-Schwinger (LS) equation \cite{hodgson1963, negele2006advances}
\begin{equation}\label{generalscatteq}
T = V \left( 1 + G_0 (E) T \right)\, , 
\end{equation}
where the operator $V$ represents the external interaction which, if we assume only two-body forces, is given by the sum over all the target nucleons  of two-body potentials describing the interaction of each target nucleon with the incident proton and $G_0 (E)$ is the free Green's function for the $(A+1)$-nucleon system.

As a standard procedure, Eq.~(\ref{generalscatteq})
is separated into a set of two coupled integral equations: the first one for the so-called $T$ matrix
\begin{equation}\label{firsttamp}
T = U \left(1+ G_0 (E) P T\right) \, 
\end{equation}
and the second one for the optical potential $U$
\begin{equation}\label{optpoteq}
U = V \left(1+ G_0 (E) Q U \right)\, .
\end{equation}
In Eqs. (\ref{firsttamp}) and (\ref{optpoteq}), the operator $P$ projects onto the elastic channel and the projection operator $Q$ is defined, as usual,  by the completeness relation $P + Q = \dblone$.

In order to develop a consistent framework to compute the optical potential $U$ and the 
transition amplitude for the elastic $NA$ scattering observables, we follow the path initiated by Kerman {\it et al.} \cite{Kerman1959551}, and subsequently improved by Picklesimer {\it et al.} \cite{PhysRevC.30.1861}, that is based on the multiple scattering theory and we retain only the first-order term, corresponding to the single-scattering approximation, where only one target-nucleon interacts with the projectile. In addition, we adopt the impulse approximation, where nuclear binding forces on the interacting target nucleon are neglected.
For all relevant details and an exhaustive bibliography we refer the reader to Ref.~\cite{PhysRevC.93.034619}, where 
the theoretical framework of the present work 
has been extensively described. 

After some lenghty manipulations, the optical potential is obtained in a factorized form (in the so called optimum factorization approximation) as the product of the free
$NN$ $t$ matrix and the nuclear matter densities
\begin{equation}
\label{optimumfact}
U ({\bm q},{\bm K};\omega) = \frac{A-1}{A} \, \eta ({\bm q},{\bm K}) 
 \sum_{N = n,p} t_{pN} \left({\bm q}, {\bm K}, \omega \right) \, \rho_N (q) \, ,
\end{equation}
where ${\bm q}$ and ${\bm K}$ are the momentum transfer and the total 
momentum, respectively, in the $NA$ reference frame, 
$t_{pN}$ represents the proton-proton ({\it pp}) and proton-neutron ({\it pn}) 
$t$ matrix, $\rho_N$ represents the neutron and proton profile density,
and $\eta ({\bm q},{\bm K})$ is the M\o ller factor, that imposes the 
Lorentz invariance of the flux when we pass from the $NA$ to the $NN$
frame in which the $t$ matrices are evaluated.  
Through the dependence of $\eta$ and $t_{pN}$ upon ${\bm K}$, the optimally factorized optical potential given in Eq.~(\ref{optimumfact}) exhibits nonlocality and off-shell effects (see Ref.~\cite{PhysRevC.93.034619}). 
The energy $\omega$ at which the matrices 
$t_{pN}$ are evaluated is fixed at one half of the kinetic
energy of the projectile in the laboratory system.

The optimally factorized optical potential is then written 
exploiting its spin-dependent component (see Sec. IIC of Ref.~\cite{PhysRevC.93.034619}) and then expanded 
on its partial-wave components. Once the $LJ$ components of the elastic transition operator are determined,
the calculation of the three scattering observables (the unpolarized differential cross section ${\rm d}\sigma/{\rm d}\Omega$, the analyzing power $A_y$, and the spin rotation $Q$) is straightforward.

Two basic ingredients are required to calculate the optical potential: 
the $NN$ potential and the neutron and proton densities of the target nucleus. For the latter quantities we follow the same path initiated in Ref. \cite{PhysRevC.93.034619} using a Relativistic Mean-Field (RMF) description \cite{Niksic20141808}. In the last years this approach has been very 
successful into the description of ground state and excited state properties of finite nuclei, in particular in a Density Dependent Meson Exchange (DDME) version, where the couplings between mesonic and baryonic fields are assumed as functions of the density itself \cite{PhysRevC.66.024306}. 
We are aware that a phenomenological description of the target is not fully consistent with the goal of a microscopic description of elastic $NA$ scattering. A forthcoming paper will be devoted to the inclusion of matter densities from ab-initio calculations. 

For the $NN$ interaction we use here two different versions of the chiral potentials at fifth order (N$^4$LO) recently derived by Epelbaum, Krebs and Mei\ss ner (EKM) \cite{Epelbaum:2014sza,Epelbaum:2014efa} and Entem, Machleidt and Nosyk (EMN) \cite{Entem:2014msa,Entem:2017gor}. 
Some basic features of these chiral potentials are outlined in the following Sec.~\ref{sec_pot}.


\subsection{N$^4$LO CHIRAL POTENTIALS}
\label{sec_pot}

Chiral Perturbation Theory (ChPT) is a perturbative technique for the description of hadron scattering amplitudes
based on expansions in powers of a parameter that can be generally defined as $(p,m_\pi)/\Lambda_b$, where $p$ is the magnitude of 3-momenta of the external particles, $m_\pi$ is the pion mass, and the symmetry breaking scale $\Lambda_b$ can be safely estimated for chiral symmetry as follows $\Lambda_b \sim 4 \pi f_\pi$ 
\cite{Scherer:2002tk} or, alternatively, using the lightest non-Goldstone meson mass as an energy scale, 
$\Lambda_b \sim m_\rho$. 

As an Effective Field Theory (EFT) \cite{Georgi:1994qn}, ChPT respects the low-energy 
symmetries of Quantum ChromoDynamics (QCD) and, up to a certain extent, is model independent and 
systematically improvable by an order-by-order expansion, 
with controlled uncertainties from neglected higher-order terms.

Nevertheless, calculations in the $NN$ sector are particularly complicated 
due to large scattering lengths and, in particular, the shallow deuteron bound state: a clear 
indication of a non-perturbative character of the $NN$ system \cite{Kaplan:1998we, VanKolck1999337}. 

At the beginning of the nineties, Steven Weinberg \cite{Weinberg:1990rz} 
proposed a practical method to calculate the $NN$ scattering amplitude:
as a first step, a nuclear potential $V$ is calculated as the sum of all irreducible
diagrams; then, solving the LS equation, $V$ is going to be iterated to all orders.

Of course the LS equation is divergent and needs to be regularized. 
In conventional field theories, the integrals are regulated and the dependence 
on the regularization parameters (cutoffs) is removed
by renormalization. At the end of the procedure the calculations
do not depend on cutoffs or renormalization scales.
A successful renormalization procedure for the $NN$ potential in which the cutoff parameter is carried to infinity is only available at leading order (LO) as it has been proven by Nogga in Ref. \cite{Nogga05}.
An extension to higher orders is, at the moment, impracticable because no reliable power counting scheme would be available \cite{EGcutoff09,ZMcutoff12}. For our purposes, cutoffs should be limited to a specific  energy domain $\Lambda\lesssim\Lambda_b$. In fact, in EFTs a different approach is pursued with the goal to maintain a regulator independent procedure (within a range of validity determined by the breakdown scale) and, at the same time, a practical power counting scheme: EFTs are usually renormalized order by order \cite{Lepage:1997cs}.

A standard choice is to multiply
the potential $V$ with a regulating function in the momentum space
\begin{equation}
\label{mom_reg}
f_\Lambda(k^\prime,k) = \exp \left( -\left(\frac{k^\prime}{\Lambda}\right)^{2m} - 
\left(\frac{k}{\Lambda}\right)^{2m} \right) \;.
\end{equation}
In general, the cutoff parameter is estimated by choosing a value for 
$\Lambda$ close to 500 MeV, safely below the EFT breakdown scale 
$\Lambda_b$. Concerning the exponents, $m=2$ or $3$ is a commonly adopted 
choice in the existing literature \cite{Machleidt:1989tm}.

At the same time, an implicit renormalization of the $NN$
amplitude is achieved by fitting to experimental phase shifts \cite{nnonline} the Low-Energy Constants (LECs) related to the contact interaction terms in the Lagrangian \cite{Epelbaum:2008ga, Machleidt:2011zz}. 

In our previous work \cite{PhysRevC.93.034619}, where we introduced our model for the first time,
calculations were performed using two different versions of the chiral potential at fourth order (\n3lo)
based on the works of Entem and Machleidt (EM) ~\cite{chiralmachleidt_n3lo} and Epelbaum Gl\"ockle, and Mei\ss ner (EGM) \cite{chiralepelbaum_n3lo}. 
Both versions employed a regulator function $f_\Lambda$ (with three choices of the cutoff: $\Lambda=$ 450, 600, and 500 (EM) or 550 (EGM) MeV) to regulate the high-momentum components in the LS equation, but they approached differently the treatment of the short-range part of the two-pion exchange (2PE) contribution, that has unphysically strong attraction.
EM treated divergent terms in the 2PE contributions with dimensional regularization (DR), while EGM used a spectral function regularization (SFR), which introduces an additional cutoff $\tilde{\Lambda}$ in the evaluation of 
the potential and, as a consequence,  also into the perturbative resummation.

Several issues arise with the SFR procedure, as pointed out in Ref. \cite{Epelbaum:2014efa}:
\begin{enumerate}
\item The inconsistency with available calculations of the three-body forces (3NF) at and 
beyond the N$^3$LO level that employ the standard DR \cite{3NF_1,3NF_2,3NF_3,3NF_4} is one of the most relevant.
As discussed in Ref. \cite{Epelbaum:2014efa}, the 
introduction of SFR on some of the 3NF
contributions, such as the ring diagrams,
appears to be a difficult task. 
\item
The values of some pion-nucleon ($\pi N$) low-energy constants, in particular the $c_i$'s, is another matter of concern. In fact, they are involved both in the $NN$ sector, through the 2PE potential, and in the long- and intermediate-range 3NFs. In Ref. \cite{chiralepelbaum_n3lo}, for example, 
the value of $c_3$ was reduced in order to tame the unphysical attraction leading to 
unphysical deeply bound states in the $NN$ system.  
\item
In EFTs it is a common procedure to estimate errors due to
truncation of the expansion at a given order by means of a cutoff dependence. Introducing $\tilde{\Lambda}$
undermines a reliable assessment of the theoretical accuracy. 
\end{enumerate}
Because of the above mentioned arguments, the authors of Ref. \cite{Epelbaum:2014sza,Epelbaum:2014efa} claim that using DR instead of SFR would be the optimal choice to calculate the chiral $NN$ potential.

Furthermore, the same authors \cite{Epelbaum:2014sza,Epelbaum:2014efa} argued that even the choice to employ a nonlocal momentum-space regulator in the $NN$ potentials \cite{chiralmachleidt_n3lo,chiralepelbaum_n3lo} leads to some  inconsistencies, considering that it affects the long-range part of the interaction, as extensively discussed in Refs. \cite{Epelbaum:2014efa,cutoff_problems_1,cutoff_problems_2}.
A possible solution to reduce finite-cutoff artifacts consists in a regularization in coordinate space. As stated in Ref. \cite{Epelbaum:2014efa},
this particular choice of a coordinate space regulator
makes the adoption of SFR for the treatment of pion exchange contributions unnecessary.
This choice would also allow one to avoid any fine-tuning 
of the low-energy constants $c_i$ and $d_i$ determined from pion-nucleon 
scattering. Such regularization has been initially adopted by Gezerlis {\it et al.} in the
construction of local
chiral $NN$ potentials up to N$^2$LO \cite{Gezerlis1,Gezerlis2}.

\subsubsection{The EKM approach}

The strategy followed in Ref. \cite{Epelbaum:2014sza,Epelbaum:2014efa} consists in a regularization for the long-range
contributions such as
\begin{equation}
V_{\rm long-range} ({\bm r}) \,\, \rightarrow \,\, V_{\rm long-range}^{\rm reg} ({\bm r}) = V_{\rm long-range} ({\bm r}) f\left(\frac{r}{R}
\right) \; ,
\end{equation}
where $f$ is a regulator function defined as
\begin{equation}
f\left(\frac{r}{R} \right) = \left(1 -\exp\left( -\frac{r^2}{R^2}\right) \right)^n \;, 
\end{equation}
and a conventional momentum space regularization, see Eq. (\ref{mom_reg}), for the contact terms with $\Lambda = 2R^{-1}$ and $m=2$. As explained in Ref. \cite{Epelbaum:2014efa}, it is necessary to choose $n\ge4$ in order to have the correct behaviours of the 2PE contributions. To guarantee more stable results from a numerical point of view, 
$n=6$ is the adopted value. Five available choices of $R$ are available:
0.8, 0.9, 1.0, 1.1, and 1.2 fm, leading to five potentials with different $\chi^2$/datum.
As shown in Tab. 3 of Ref. \cite{Epelbaum:2014efa}, they are almost equivalent for energies below 200 MeV, with larger discrepancies for higher energies, in particular for the softest (1.2 fm) and the hardest cases (0.8 fm).   

\subsubsection{The EMN approach}

On the other hand, Machleidt {\it et al.} \cite{Entem:2014msa, Entem:2017gor} pursued a slightly more conventional approach to develop a $NN$ potential at \chiral4lo. They employed a SFR with a cutoff $\tilde{\Lambda} = 700$ MeV (while, at lower orders, $\tilde{\Lambda} = 650$ MeV) in order to regularize the loop contributions. The long-range parts are constrained by a recent Roy-Steiner (RS) analysis by Hoferichter {\it et al.} \cite{Hoferichter:2015tha, Hoferichter:2015hva}. With RS equations the LECs can be extracted from the subthreshold point in $\pi N$ scattering data with extremely low uncertainties (see Tab. II of Ref. \cite{Entem:2017gor} for more details). As a last step, to deal with infinities in the LS equation, a conventional regulator function
(\ref{mom_reg}) is employed, with $\Lambda = 450, 500$, and $550$ MeV as available choices, and $m=2$ and $4$
for multi-pion and single-pion exchange contributions, respectively. 
For all details we refer the reader to Refs. \cite{Entem:2014msa, Entem:2017gor}.
The \chiral4lo potential produced with the previous approach is able to reproduce a very large $NN$ database (see Sec.IIIA of Ref. \cite{Entem:2017gor}) with a``realistic" $\chi^2/$datum $\sim 1.15$.

It is therefore very interesting to compare these two different approaches and to study the differences produced on  elastic $NA$ scattering observables by 
the different $NN$ potentials and their regularizations. 
In particular, our goal is to study what regularization prescription is more suitable and successful in reproducing empirical data.
In the following, results are presented and compared for the $NN$ Wolfenstein amplitudes and for elastic proton-scattering observables on ${}^{12}$C, ${}^{16}$O, and ${}^{40}$Ca nuclei. 


\section{Results}
\label{sec_res}

\subsection{$NN$ AMPLITUDES}
\label{sec_amp}

In this section we present and discuss the theoretical results for the {\it pp} and {\it pn} Wolfenstein amplitudes \cite{PhysRev.85.947,NNamplitude}. 
For the $J=0^+$ nuclei we are interested in the present work, only $a$ and $c$ amplitudes survive and they are connected to the central and the spin-orbit part of the $NN$ $t$ matrix, respectively (more details can be found, e.g., in Sec. II B of Ref. \cite{PhysRevC.93.034619}).

\begin{figure}[t]
\begin{center}
\includegraphics[scale=0.5]{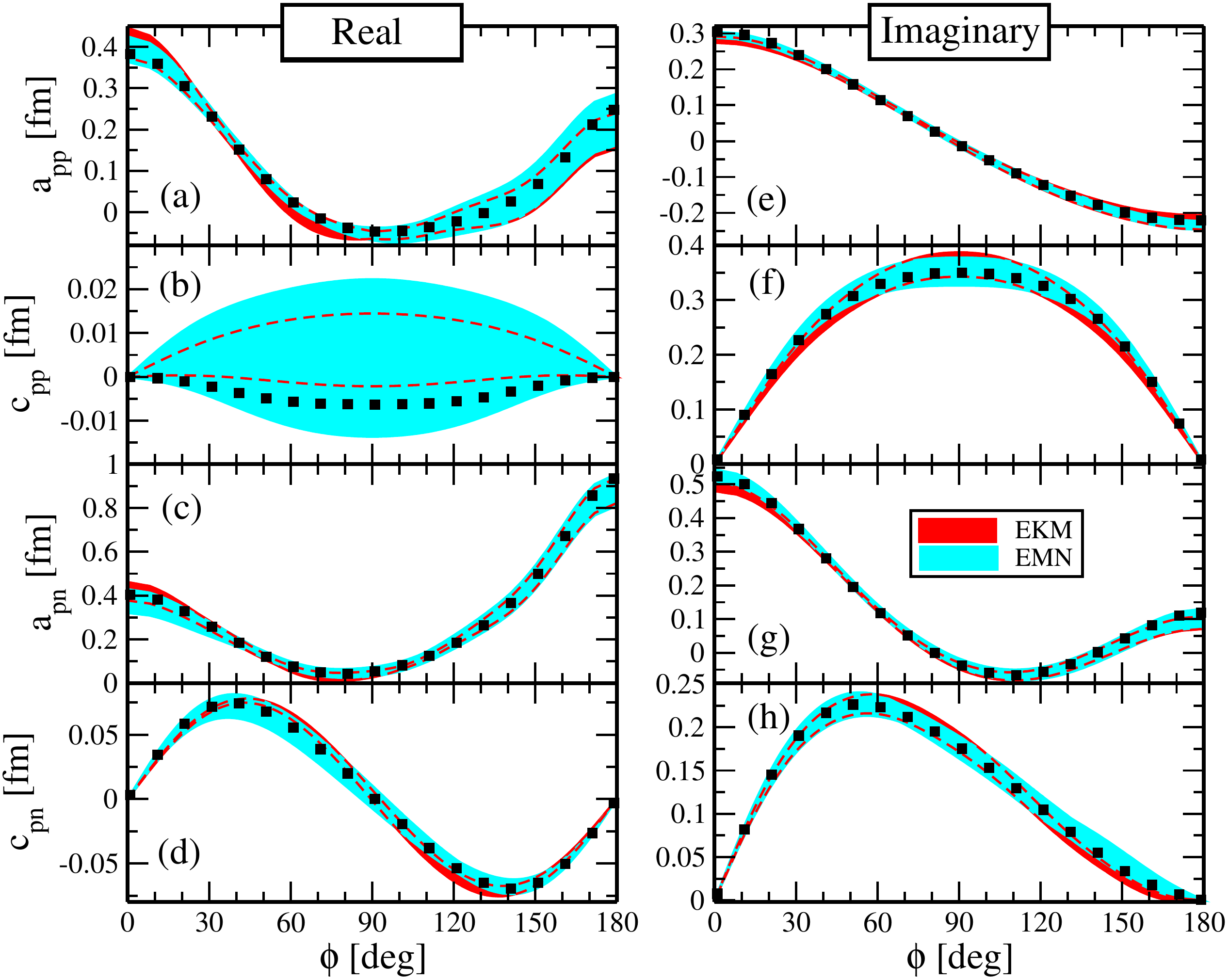}
\caption{\label{fig_nn_amp}  (Color online) Real (left panel) and imaginary (right panel) parts of {\it pp} and {\it pn} $a$ and $c$ Wolfenstein amplitudes as functions of the center-of-mass $NN$ angle $\phi$. All the amplitudes are computed at $200$ MeV using one of the EKM \cite{Epelbaum:2014sza,Epelbaum:2014efa} potentials (red bands determined by $R =0.9$ fm) and one of the EMN \cite{Entem:2014msa,Entem:2017gor} potentials (cyan bands) which uses a momentum cutoff $\Lambda = 500$ MeV. To estimate theoretical errors, we used Eq. (\ref{er_est})
with $\Lambda_b = 600$ MeV.
Empirical data are taken from Ref.~\cite{nnonline}.}
\end{center}
\end{figure}

All calculations are performed with one of the EKM \cite{Epelbaum:2014sza,Epelbaum:2014efa} potentials (red bands in Fig. \ref{fig_nn_amp}), corresponding to $R=0.9$ fm, and with the EMN \cite{Entem:2014msa, Entem:2017gor} potential (cyan bands in Fig. \ref{fig_nn_amp}) which employs a momentum cutoff regularization with $\Lambda = 500$ MeV. 

In both cases we plot bands and not just lines because, for this class of chiral potentials, it is possible to assess theoretical errors associated with the truncation of the chiral expansion. In order to estimate the size of this theoretical uncertainties, we follow the same approach proposed in Refs. \cite{Epelbaum:2014sza,Epelbaum:2014efa}.
Given an observable $\mathcal{O} (p)$ as a function of the center of mass momentum $p$, 
the uncertainty $\Delta \mathcal{O}^{\rm n} (p)$ at order $n$
is given by the size of neglected higher-order terms. For example, at \chiral4lo order we have
\begin{eqnarray}
\label{er_est}
\Delta \mathcal{O}^{N^4LO} (p) =  {\rm max} &\; & \left( Q^{6} \times \left| \mathcal{O}^{\rm LO} (p) \right|, \right. 
\nonumber\\
& ~ &  Q^{4} \times \left| \mathcal{O}^{\rm LO} (p) - \mathcal{O}^{\rm NLO} (p) \right|, 
Q^{3} \times \left| \mathcal{O}^{\rm NLO} (p) - \mathcal{O}^{\rm N^2LO} (p) \right|, \nonumber\\
& ~ &  \left. Q^{2} \times \left| \mathcal{O}^{\rm N^2LO} (p) - \mathcal{O}^{\rm N^3LO} (p) \right|, 
Q \times \left| \mathcal{O}^{\rm N^3LO} (p) - \mathcal{O}^{\rm N^4LO} (p) \right| \right) \; ,
\end{eqnarray}
where $Q$ is defined as follows
\begin{equation}  
\label{expansion}
Q =\max \left( \frac{p}{\Lambda_b}, \; \frac{M_\pi}{\Lambda_b} \right)\,,
\end{equation}
and $\Lambda_b = 600$ MeV is an optimal choice \cite{Epelbaum:2014sza,Epelbaum:2014efa,Mac_mail}.
Concerning error estimates, other prescriptions can be used \cite{Mac_mail}.
For example, the simplest one would be to explore cutoff dependences. We have performed some preliminary calculations and, in our opinion, the method introduced in Refs. \cite{Epelbaum:2014sza,Epelbaum:2014efa} seems to
be the best choice.

We also tested that predictions based on different values of $R$ and $\Lambda_b$ are quite close and consistent with each other (as remarked in Ref. \cite{Epelbaum:2014sza} larger values of R are probably less accurate due to a larger influence of cutoff artifacts). We are therefore confident that for our present purposes showing results with only a single potential of the EKM set will not affect our conclusions in any way.
The same assumption can be made about the EMN potentials: changing the cutoffs does not lead to sizeable differences in the $\chi^2/$datum (see Tab.VIII in Ref. \cite{Entem:2017gor}) and it is safe to perform calculations
with only a single potential.

In Fig.~\ref{fig_nn_amp} the theoretical results for the real and imaginary parts of the {\it pp} and {\it pn} 
amplitudes ($a$ and $c$), computed at an energy of $200$ MeV, are shown as functions of the 
center-of-mass $NN$ angle $\phi$ and compared with the experimental amplitudes, which have been extracted from the experimental $NN$ phase shifts \cite{nnonline}.
We have chosen a rather high energy for our calculations in order to enlarge the differences among the potentials employed.
As shown in Figs. 1 and 2 of Ref. \cite{PhysRevC.93.034619}, no appreciable differences are given by different $NN$ potentials at lower energies. 
In Fig.~\ref{fig_nn_amp} the experimental data are globally very well reproduced by the theoretical results, with the only remarkable exception of the real part of the $c_{pp}$ amplitude that is overestimated.
It must be considered, however, that $c_{pp}$ is a very small quantity, i.e., two orders of magnitude smaller than the respective imaginary part, and it will only provide a very small contribution to the optical potential. 
We do not find appreciable differences with respect to the choice of the $NN$ potential,
in fact the cyan bands largely overlap the red bands for any amplitudes. In both cases,
the bands are very narrow, maybe with mild exceptions for the real part of $a_{pp}$ 
and the imaginary components of $c_{pp}$ and $c_{pn}$. As a consequence, we can conclude that the $NN$ sector has already reached a robust convergence at \chiral4lo and we do not expect large contributions from the N$^5$LO extension \cite{n5lo_1, n5lo_2}.
 

\subsection{ELASTIC PROTON-NUCLEUS SCATTERING OBSERVABLES}

In this section we present and discuss our numerical results for the  $pA$ elastic scattering observables calculated with the microscopic
optical potential obtained within the theoretical framework described in the  previous sections. 
We consider elastic proton scattering on $^{12}$C, $^{16}$O, and $^{40}$Ca.

The main goal of our work is to investigate the sensitivity of the results to 
the choice of the $NN$ potential and to assess theoretical uncertainties for 
the scattering observables. 
In Ref. \cite{PhysRevC.93.034619} we studied the limits of applicability of chiral potentials in terms of the proton energy.
In the present work we show results for a single proton energy of $200$ MeV,  
a value that represents a good compromise between the limits of applicability 
of our model (the results shown in Ref. \cite{PhysRevC.93.034619} indicate that for energies larger than $200$ MeV the agreement between the results from chiral potentials and data gets worse and it is plausible to believe that ChPT is no longer applicable) and the necessity to emphasize the differences between the 
$NN$ potentials employed, that increase with increasing energy.

\begin{figure}[t]
\begin{center}
\includegraphics[scale=0.5]{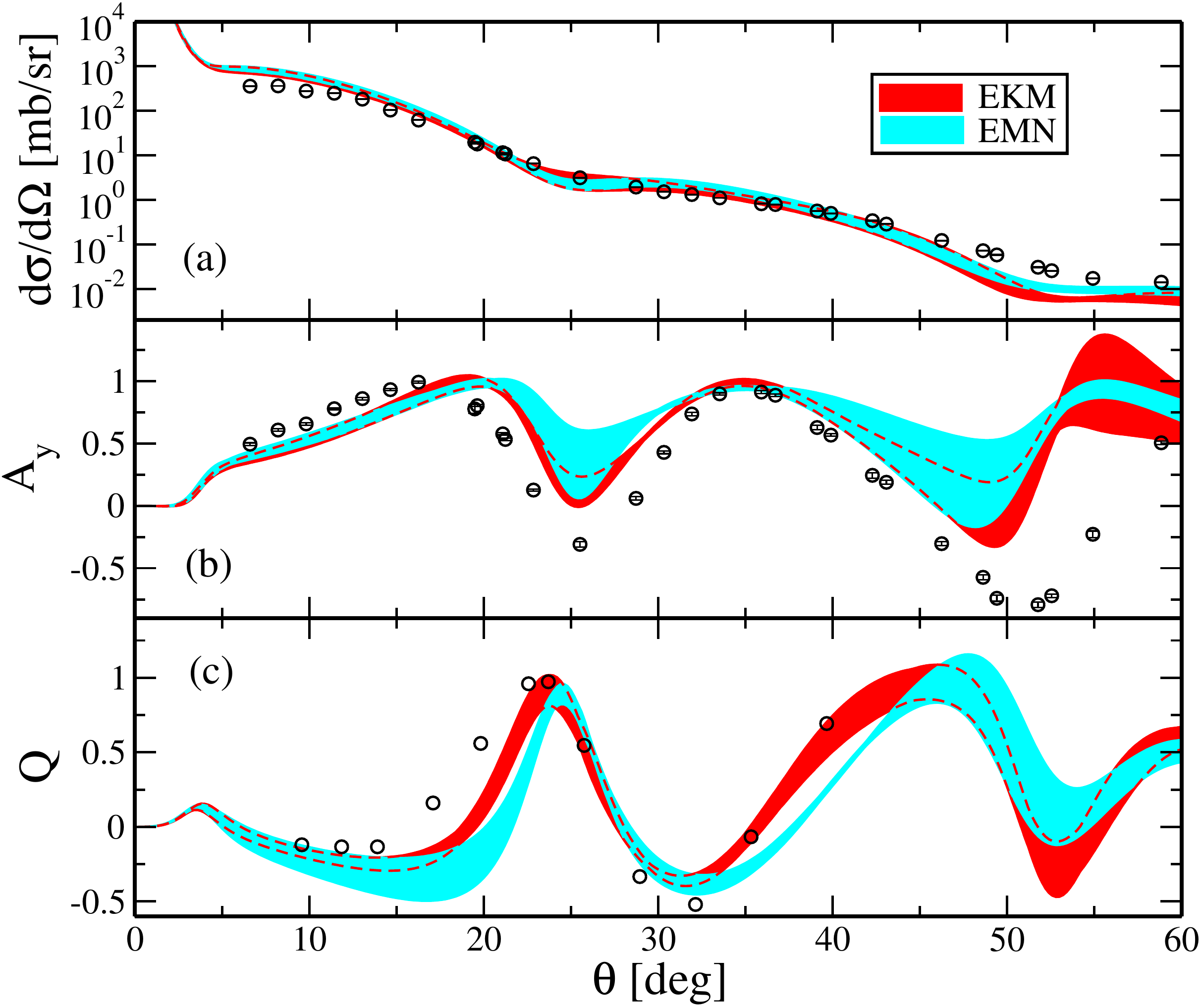}
\caption{\label{fig_o16_200} (Color online) 
Scattering observables (differential cross section $d\sigma/d\Omega$, 
analyzing power $A_y$, and spin rotation $Q$) as a function of the center-of-mass scattering angle $\theta$  
for elastic proton scattering on ${}^{16}$O computed at $200$ MeV
(laboratory energy). We employ one of the EKM \cite{Epelbaum:2014sza,Epelbaum:2014efa} potentials (red bands determined by $R =0.9$ fm) and one of the EMN \cite{Entem:2014msa,Entem:2017gor} potentials (cyan bands) which uses a momentum cutoff $\Lambda = 500$ MeV. 
To estimate theoretical errors, we used Eq. (\ref{er_est})
with $\Lambda_b = 600$ MeV.
Coulomb distortion is included as explained in Ref. \cite{PhysRevC.93.034619}.
Empirical data are taken from Refs.~\cite{kelly, exfor}.}
\end{center}
\end{figure}

\begin{figure}[t]
\begin{center}
\includegraphics[scale=0.5]{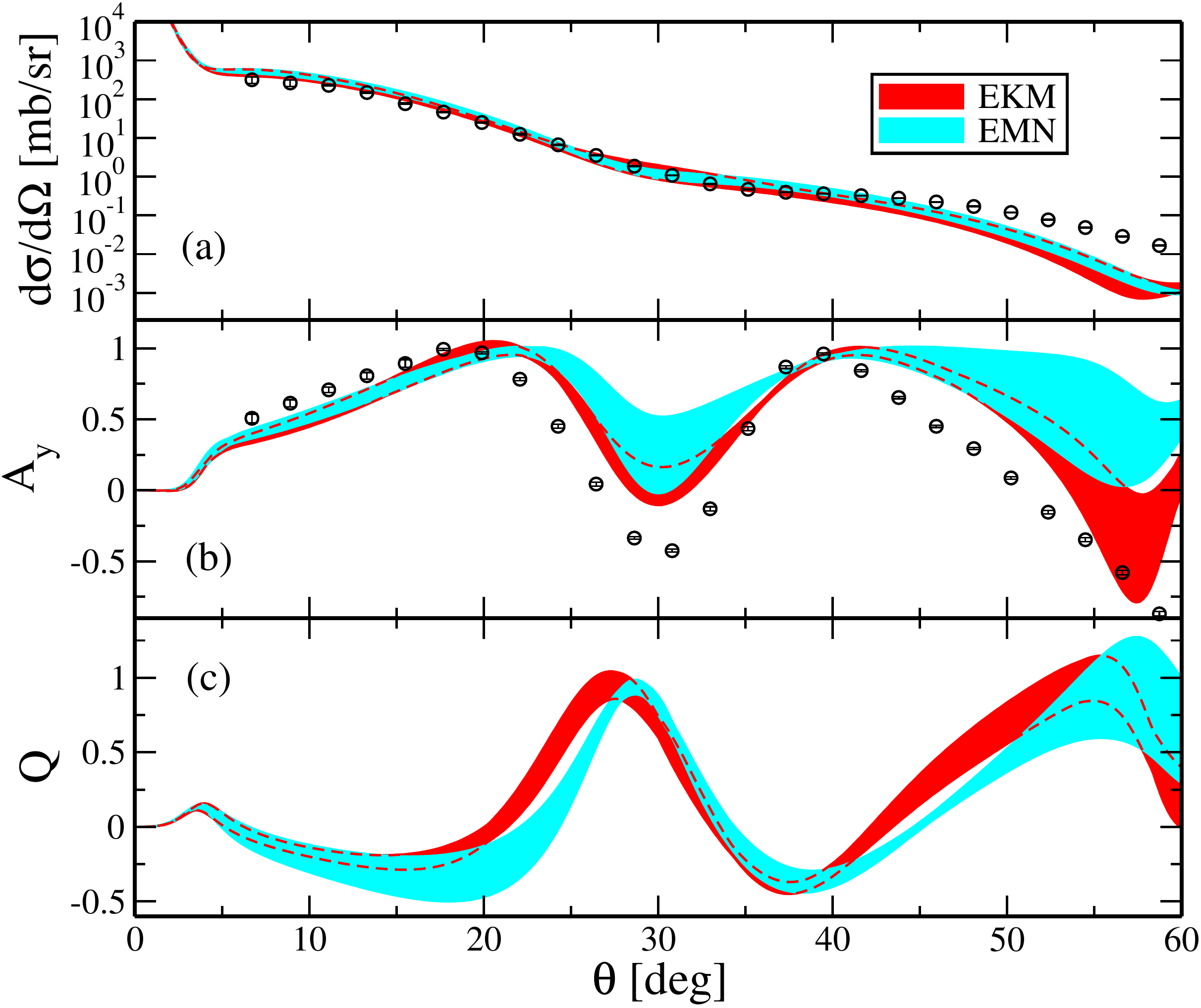}
\caption{\label{fig_c12_200} (Color online) The same as is Fig. \ref{fig_o16_200} for ${}^{12}$C
at an energy of $200$ MeV. Empirical data are taken from Refs. \cite{kelly, exfor}. }
\end{center}
\end{figure}

\begin{figure}[t]
\begin{center}
\includegraphics[scale=0.5]{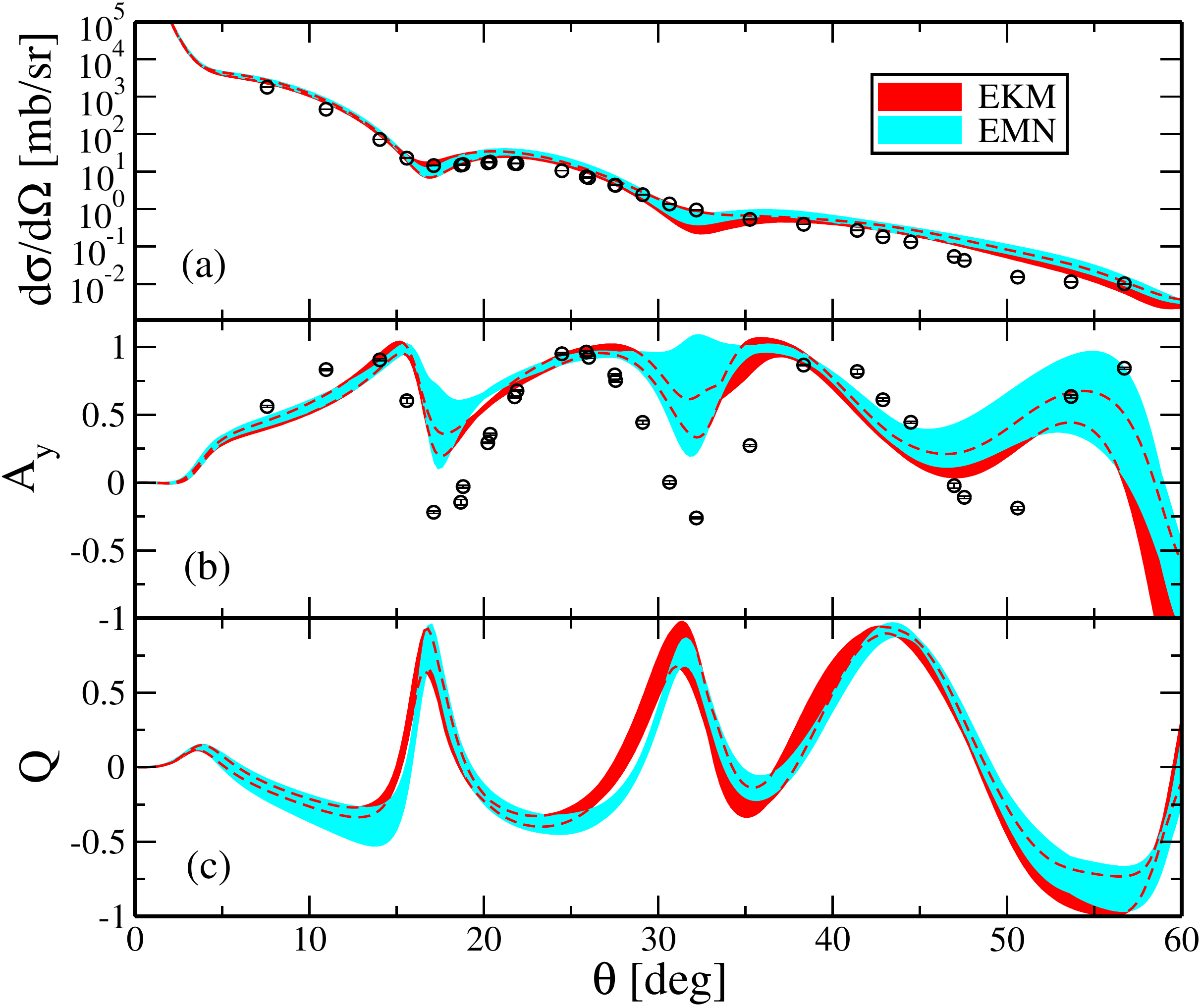}
\caption{\label{fig_ca40_200} (Color online) The same as is Fig. \ref{fig_o16_200} for ${}^{40}$Ca
at an energy of $200$ MeV. Empirical data are taken from Refs. \cite{kelly,exfor}.   }
\end{center}
\end{figure}

In Figs.~\ref{fig_o16_200}, \ref{fig_c12_200}, and \ref{fig_ca40_200} we show the differential cross section 
(${\rm d}\sigma/{\rm d}\Omega$), the analyzing power $A_y$, and the spin rotation $Q$ for elastic proton scattering 
on $^{16}$O, $^{12}$C, and $^{40}$Ca, respectively, as functions of the center-of-mass scattering angle $\theta$. The results are compared with the experimental data taken from Refs.~\cite{kelly, exfor}.

As in Sec. \ref{sec_amp}, all calculations are performed with one of the EKM potentials ($R=0.9$ fm) and with one of the EMN potentials (with $\Lambda = 500$ MeV). Red and cyan bands for the EKM and EMN results are produced following the above mentioned prescription, see Eq. (\ref{er_est}), with $\Lambda_b = 600$ MeV.
The Coulomb interaction between the proton and the target nucleus is included in the calculations as described in Ref. \cite{PhysRevC.93.034619}.

The first nucleus we consider is $^{16}$O, in Fig.~\ref{fig_o16_200}, that has been also investigated in Ref. \cite{PhysRevC.93.034619}. 
At  the calculated energy of $200$ MeV all sets of potentials, regardless of their theoretical differences, give very similar results for the differential cross section. Small discrepancies in comparison with empirical data appear at small ($\theta \le 5^o$) and large ($\theta \ge 50^o$) angles, but the experimental cross section is well reproduced by all potentials in the minimum region, between $20$ and $25$ degrees.
Concerning the analyzing power $A_y$, both potentials overestimate the experimental data for angles larger than 20 degrees but the overall behaviour 
is nicely reproduced. 
The numerical results for the spin rotation $Q$ exhibit a good agreement with empirical data. This is a non trivial task considering that 
polarization observables are usually more difficult to reproduce.
The cyan and red bands, assessing theoretical errors due to the truncation of the chiral expansion, for both potentials are narrow at small angles and a bit larger around the minima and at larger angles, where theoretical uncertainties increase and also the agreement with data declines.

In comparison with the corresponding results in Fig.~8 of Ref. \cite{PhysRevC.93.034619}, which are calculated for the same nucleus at the same energy and within the same theoretical framework for the $NA$ optical potential, but with the EM and EGM chiral potentials at fourth order (N$^3$LO), the present results in Fig.~\ref{fig_o16_200} give a comparable, and in general not particularly better, description of the experimental data.
From this point of view, they confirm our previous results of Ref. \cite{PhysRevC.93.034619}. The aim of our investigation was not to obtain 
a perfect agreement with the data (although not perfect, the agreement can be considered reasonable if we bear in mind the approximations of our model), but to study the applicability of microscopic two-body chiral potentials in the construction of an optical potential. More specifically, in this work, our aim is to check the convergence of the ChPT perturbative expansion and the sensitivity of the results to the choice of the $NN$ potential and to the adopted regularization prescription.  
Different $NN$ potentials, able to give equivalently good descriptions of $NN$ elastic-scattering data, may have a different off-shell behaviour, and it is this behaviour, that cannot be tested in the comparison with $NN$ scattering data, that can produce different results when the $NN$ potentials  are used to calculate the optical potential for elastic $NA$ scattering.     

Also for $^{12}$C in Fig.~\ref{fig_c12_200} all sets of $NN$ potentials give very close results for the calculated differential cross sections and somewhat larger, although not crucial, differences for the analyzing power $A_y$ and the spin rotation $Q$. The experimental cross section is well described by our results for angles up to $\theta \simeq 45^o$ and somewhat underestimated at larger angles. Our calculations are able to describe the behaviour (the shape better than the size) of the experimental $A_y$. No empirical data are available for $Q$.

For $^{40}$Ca in Fig.~\ref{fig_ca40_200} all sets of $NN$ potentials give very close results and a generally good description of the experimental cross section. The experimental analyzing power $A_y$ is somewhat overestimated (but for small angles), in particular around the minima.

\begin{figure}[t]
\begin{center}
\includegraphics[scale=0.5]{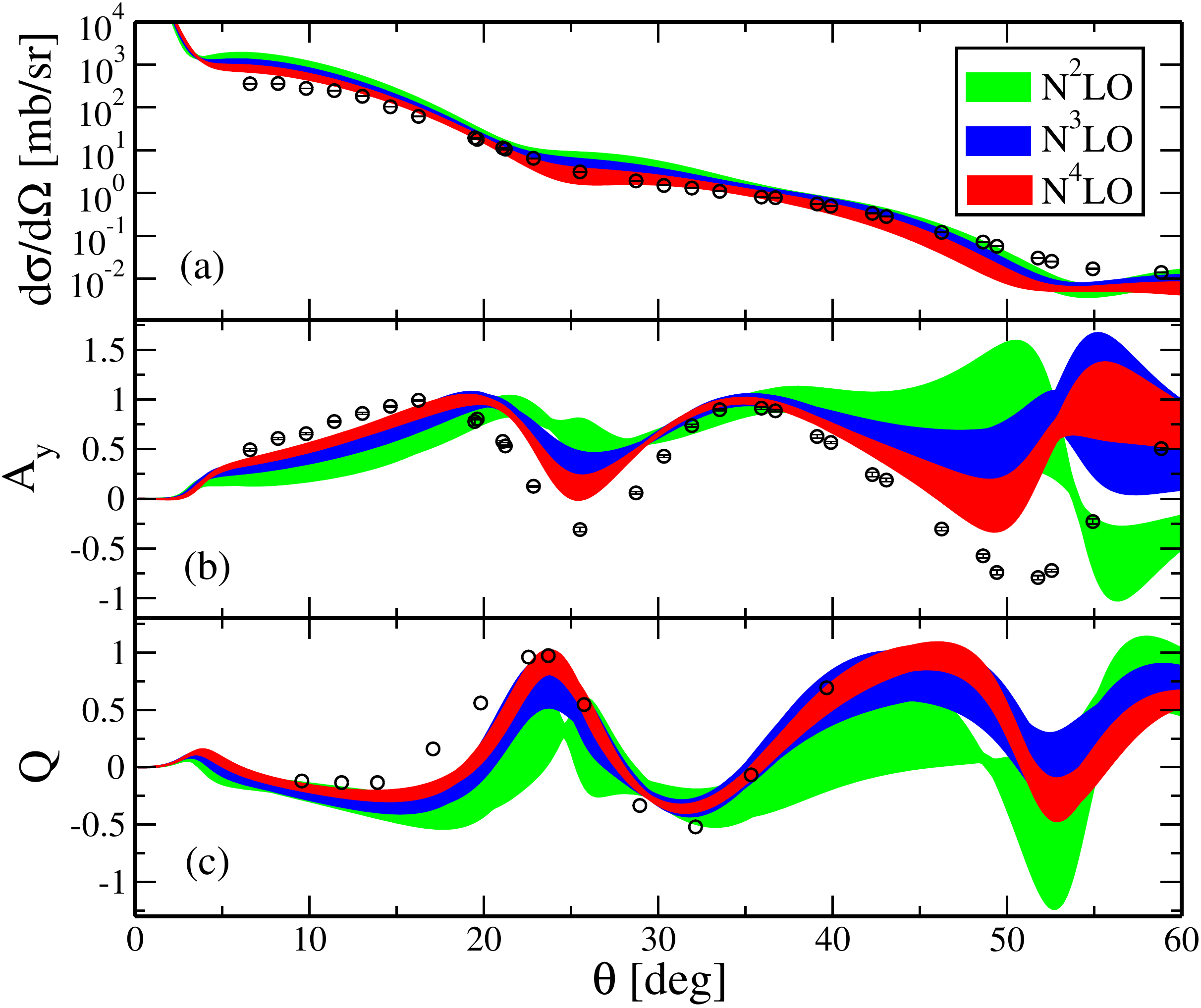}
\caption{\label{fig_o16_200_orders} (Color online)
Scattering observables as a function of the center-of-mass scattering angle $\theta$  
for elastic proton scattering on ${}^{16}$O computed at $200$ MeV
(laboratory energy) with the EKM potential \cite{Epelbaum:2014sza,Epelbaum:2014efa} at different orders: 
green bands are the N$^2$LO results, blue and red bands are the N$^3$LO and N$^4$LO
results, respectively. Empirical data are taken from Refs.~\cite{kelly,exfor}.}
\end{center}
\end{figure}

Generally speaking, red bands are narrower than cyan ones, suggesting a stronger control of theoretical errors 
at N$^4$LO for the EKM potentials. Concerning the order by order convergence pattern 
(N$^2$LO, N$^3$LO, N$^4$LO) for the scattering observables of elastic proton scattering on $^{16}$O, an example 
calculated with the EKM potential is presented Fig.~\ref{fig_o16_200_orders}. The error bands and therefore the theoretical uncertainties are clearly reduced from N$^2$LO to N$^4$LO, the convergence pattern is clear, and we can conclude that convergence  has been reached at N$^4$LO. We do not expect large contributions from the higher-order extensions in the $NN$ sector, but it could  be interesting to see what happens with $NN$ potentials at 
N$^5$LO \cite{n5lo_1, n5lo_2}. 

\section{Conclusions}
\label{sec_concl}

In a previous paper~\cite{PhysRevC.93.034619} we derived a new microscopic optical potential for elastic $pA$ scattering from $NN$ chiral potentials at fourth order (N$^3$LO)~\cite{chiralmachleidt_n3lo,chiralepelbaum_n3lo}, with the purpose to study the domain of applicability of microscopic two-body chiral potentials in the construction of an optical potential. 
In the present work a microscopic optical potential has been derived, within the same theoretical framework and adopting the same approximations as in Ref.~\cite{PhysRevC.93.034619}, from $NN$ chiral potentials at fifth order (N$^4$LO) based on the recent works of Epelbaum, Krebs and 
Mei\ss ner~\cite{Epelbaum:2014sza,Epelbaum:2014efa} and Entem, Machleidt and Nosyk~\cite{Entem:2014msa, Entem:2017gor}. Our main aims were to check the convergence of the ChPT perturbative expansion, assessing theoretical errors associated with the truncation of the chiral expansion, and to compare the results produced by the different $NN$ chiral potentials and their different regularizations on elastic $NA$ scattering observables. 

Numerical results have been presented for the $pp$ and $np$ Wolfenstein amplitudes ($a$ and $c$), that are employed in the calculation of the optical potential to compute the $NN$  $t$ matrix, and for the  observables (the unpolarized differential cross section ${\rm d}\sigma/{\rm d}\Omega$, the analyzing power $A_y$, and the spin rotation $Q$) of elastic proton scattering from $^{12}$C, $^{16}$O, and $^{40}$Ca nuclei.  
A single proton energy of $200$ MeV has been chosen for all the calculations. The chosen energy value is rather high, in order to enlarge the differences between the different potentials, that increase with increasing energy, but within the limit of applicability for chiral potentials.  It was indeed shown in Ref.~\cite{PhysRevC.93.034619} that for energies larger than $200$ MeV the agreement between the results from chiral potentials and data gets worse and it is plausible to believe that ChPT is no longer applicable. 

The experimental $pp$ and $np$ $a$ and $c$ amplitudes are globally very well reproduced by both $NN$ chiral potentials, with the only exception of the real part of the $c_{pp}$ amplitude, which is anyhow extremely small and provides a practically  negligible contribution to the optical potential. Theoretical errors associated with the truncation of the chiral expansion are generally very small, indicating that a robust convergence has already been reached at N$^4$LO. The results for elastic $pA$ scattering observables show that the different chiral potentials give, for all three nuclei considered, very similar cross sections, in a generally good agreement with the experimental data. Polarization observables are more sensitive to the differences in the $NN$ interaction. For $^{16}$O the numerical results, in particular with the EKM potential, are in fair agreement with the experimental spin rotation (empirical data are not available for $^{12}$C and $^{40}$Ca).  
For all three nuclei both EKM and EMN potentials describe the overall behaviour of the experimental analyzing power but the size is somewhat overestimated at larger scattering angles.

The bands associated with the theoretical errors due to the truncation of the chiral expansion are small for the cross sections and larger for the polarization observables. The bands are somewhat larger for the EMN potential, suggesting a stronger control of theoretical errors at N$^4$LO for the EKM potential.
The order by order convergence pattern (an example has been presented for $^{16}$O with the EKM potential) is clear and we can conclude that  convergence has been reached at N$^4$LO and we do not expect large contributions from the higher-order extensions in the $NN$ sector. Anyhow, it will be interesting to discuss in a forthcoming paper the results with $NN$ potentials at N$^5$LO \cite{n5lo_1, n5lo_2}. 

The agreement of the present results with empirical data is comparable with (but in general not better than) the agreement obtained in Ref.~\cite{PhysRevC.93.034619} with chiral potentials at fourth order (N$^3$LO). A better agreement would require improving or reducing the approximations adopted in the calculation of the optical potential. 
As possible improvements, in the future we plan to include three-body forces and nuclear-medium effects and to go beyond the optimum factorization approximation and calculate the optical potential from a full-folding integral.

In addition, we plan to extend our investigation to  $N \ne Z$ nuclei. In particular for these nuclei, proton and neutron densities from ab-initio calculations would improve the microscopic character and the predictive power of the optical potential.

\section{Acknowledgements}
The authors are deeply 
grateful to E. Epelbaum (Institut f\"ur Theoretische Physik II,
Ruhr-Universit\"at Bochum) for providing the chiral potential of Ref. \cite{Epelbaum:2014sza,Epelbaum:2014efa}
and R. Machleidt (Department of Physics, University of Moscow, Idaho) for the chiral potential of 
Ref. \cite{Entem:2014msa,Entem:2017gor}.

\bibliography{biblio}

\end{document}